\documentstyle[aps,prb,epsf,multicol]{revtex}
\tighten
\begin{document}
\draft
\newcommand{\bn}{{\bf n}}
\newcommand{\bp}{{\bf p}}
\newcommand{\br}{{\bf r}}
\newcommand{\bq}{{\bf q}}
\newcommand{\bj}{{\bf j}}
\newcommand{\bE}{{\bf E}}
\newcommand{\eps}{\varepsilon}
\newcommand{\la}{\langle}
\newcommand{\ra}{\rangle}
\newcommand{\cK}{{\cal K}}
\newcommand{\cD}{{\cal D}}
\newcommand{\hp}{\hat p}
\newcommand{\hq}{\hat q}
\newcommand{\hx}{\hat x}
\newcommand{\hH}{{\hat H}_0}
\newcommand{\mybeginwide}{
    \end{multicols}\widetext
    \vspace*{-0.2truein}\noindent
    \hrulefill\hspace*{3.6truein}
}
\newcommand{\myendwide}{
    \hspace*{3.6truein}\noindent\hrulefill
    \begin{multicols}{2}\narrowtext\noindent
}
\title{Charge Densities and Charge Noise in Mesoscopic Conductors}
\author{M.\ B\"uttiker}
\address{Department of Theoretical Physics, 
University of Geneva, CH-1211 Geneva 4, 
Switzerland}
\date\today
\maketitle
\bigskip
\begin{abstract}
We introduce a hierarchy 
of density of states
to characterize the charge distribution 
in a mesoscopic conductor.
At the bottom of this hierarchy are the 
partial density of states which represent the contribution 
to the local density of states if both the incident and 
the out-going scattering channel is prescribed. 
The partial density of states play a prominent role 
in measurements with a scanning tunneling microscope 
on multiprobe conductors in the presence of current flow.
The partial density of states determine the degree of dephasing 
generated by a weakly coupled voltage probe. In addition 
the partial density of states determine the frequency-dependent 
response of mesoscopic conductors in the presence of slowly oscillating 
voltages applied to the contacts of the sample. 
The partial density of states permit the formulation of a Friedel sum 
rule which can be applied locally. 
We introduce the off-diagonal elements of the partial density of 
states matrix to describe charge fluctuation processes. 
This generalization leads to a local Wigner-Smith life-time matrix.

\end{abstract}

\begin{multicols}{2}
\narrowtext
The characterization of the charge distribution and its fluctuations
are a central aspect of electrical conduction theory. In this
work we are concerned with mesoscopic conductors, structures which are 
so small and cooled to such low temperatures that the wave nature 
of electron motion becomes apparent. Three recent reviews provide
an entry to the literature of this very active field \cite{ABG,GH,BB}.
Of interest is the charge distribution 
in such a conductor and its fluctuations under equilibrium and 
non-equilibrium conditions.
The charge distribution is an intersting 
quantity in itself. However, in addition, it is important for the solution 
of a number of problems. We mention here only a few: Already, in the 
equilibrium state, the charge distribution needs to be known to determine 
the effective equilibrium one-electron potential. It is this potential 
which determines the scattering properties of the conductor and
thus the conductance coefficients. If voltages are applied 
which drive the conductor away from the equilibrium state, typically, 
the charge distribution also undergoes a bias dependent modification.
Thus the charge distribution is needed to find the non-linear 
I-V-characteristic of an electrical conductor. Nanoscopic
contacts to a sample measure local density of states and 
thus a charge distribution. Knowledge of the charge 
distribution is essential for the discussion of the dynamic conductance 
like the ac-conductance or photon assisted transport. The charge 
distribution also determines the capacitance coefficients of 
various spatial regions within the mesoscopic structure and also 
vis-a-vis nearby gates. 

\section{Local partial density of states} 

It is the purpose of this work to present 
a discussion of the charge distribution \cite{BTP2,MB93} 
and the charge noise \cite{MATH}
based on the scattering approach. Similar to 
the discussion of linear transport \cite{LAND,IMRY,MB86}, 
we view the sample as a target which transmits 
or reflects electrical carriers into the contacts 
connected to the sample. A particular example \cite{GRAM} 
of a structure of 
interest here is shown in Fig. 1.  
For purely elastic 
scattering the sample is described 
by a scattering matrix ${\bf s}$. We label the different 
contacts to the sample by $\alpha = 1, 2, ...$.
Each contact
is treated as perfect wave guide which permits 
the definition of incoming and out-going 
quantum channels. 
Each contact has at a given energy a set of
quantum channels with quantum numbers labeled $m$. 
The scattering matrix element $s_{\alpha m, \beta n}$ relates the out going 
current amplitude in contact $\alpha$ in channel $m$ 
to the incoming current amplitude in contact $\beta$ in channel $n$. 
For the following discussion it is useful to introduce the 
submatrice ${\bf s}_{\alpha \beta}$ which combines 
all scattering matrix elements with the same contact indices.
This is a matrix of dimension $M_{\alpha} x M_{\beta}$
for a conductor with $M_{\alpha}$ open scattering channels 
in contact ${\alpha}$ and $M_{\beta}$ open channels 
in contact $\beta$. 
We will now show that the equilibrium and non-equilibrium 
charge distribution can be described with the help of a hierarchy of density 
of states. On top of this pyramid is the local density of states.
At the bottom of this pyramid are the 
{\it local partial density of states}\cite{BTP2,MATH,GRAM,GASP} 
\begin{equation}
\nu (\alpha , {\bf r},  \beta ) = - \frac{1}{4\pi i}
Tr \left[ {\bf s}_{\alpha \beta }^{\dagger}
\frac{\delta {\bf s}_{\alpha \beta }}{e\delta U({\bf r})} - 
\frac{\delta {\bf s}_{\alpha \beta }^{\dagger}}{e\delta U(\bf r)}
{\bf s}_{\alpha \beta }\right]\;\;
\label{lpdos2}
\end{equation}
which give the contribution to the local density of states at point ${\bf r}$
of carriers incident in contact $\beta$ 
and leaving the conductor through contact $\alpha$.
In Eq. (\ref{lpdos2}) the trace 
\begin{figure}
\narrowtext
\centerline{
\epsfxsize7cm
\epsffile{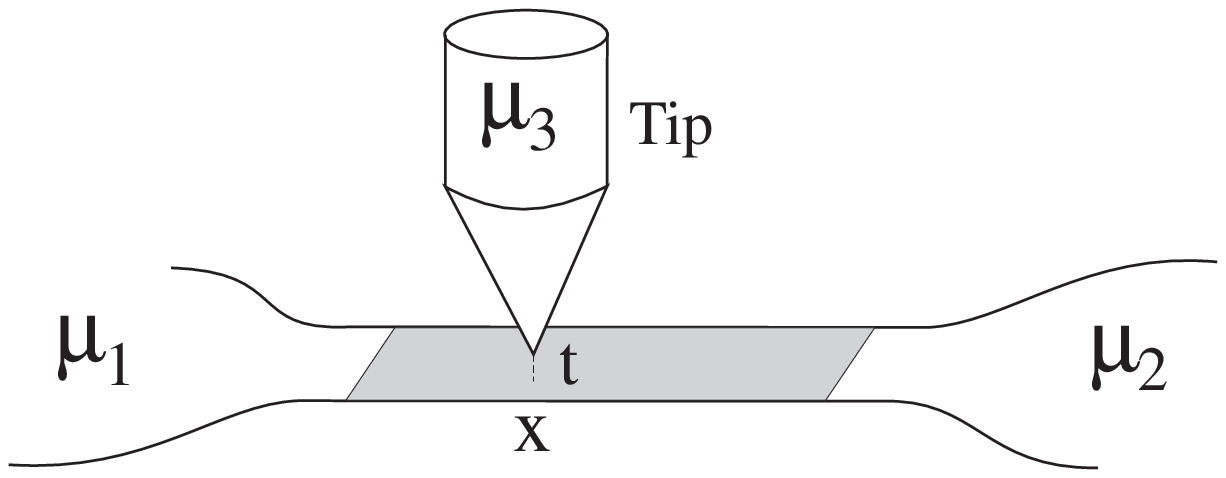}}
\vspace{0.4cm}
\caption{Mesoscopic conductor with a tunneling contact 
with coupling energy $t$ at $x$.
After Ref. \protect\onlinecite{GRAM}.}
\label{eintip}
\end{figure}
\noindent
represents a summation over channel 
indices. 
In the partial 
density of states we specify both the incident 
contact (a pre-selection) and the exiting contact (a post-selection).  
The local partial density of states is obtained by investigating
the change of the scattering matrix due to a small additional 
potential perturbation $\delta U(\bf r)$ at point ${\bf r}$. 

On the next higher level in the pyramid of density of states 
are the {\it injectivities} and {\it emissivities}. The injectivity 
of contact $\beta$ gives the contribution to the density 
of states of carriers entering the conductor through $\beta$
irrespective of the contact through which they leave the 
sample. The injectivity $\nu ({\bf r},  \beta )$ into point 
${\bf r}$ of contact $\beta$ is obtained 
by summing all partial density of states $\nu (\alpha , {\bf r},  \beta )$
over the first contact index, 
\begin{equation}
\nu ({\bf r},  \beta ) = -\frac{1}{4\pi i}
\sum_{\alpha} Tr \left[ {\bf s}_{\alpha \beta }^{\dagger}
\frac{\delta {\bf s}_{\alpha \beta }}{e\delta U({\bf r})} - 
\frac{\delta {\bf s}_{\alpha \beta }^{\dagger}}{e\delta U(\bf r)}
{\bf s}_{\alpha \beta }\right]\;\;.
\label{inject}
\end{equation} 
Using the unitarity of the scattering matrix 
we can also write for the injectivity, 
\begin{equation}
\nu ({\bf r},  \beta ) = -\frac{1}{2\pi i}
\sum_{\alpha} Tr \left[ {\bf s}_{\alpha \beta }^{\dagger}
\frac{\delta {\bf s}_{\alpha \beta }}{e\delta U({\bf r})} \right]\;\;.
\label{inject1}
\end{equation} 
The emissivity of point ${\bf r}$
into contact $\alpha$ is that portion of 
the local density of states which consists of carriers
which leave the sample through contact $\alpha$
irrespective of the contact through which they entered the sample. 
It is obtained by summing the partial density of states
over the second contact index $\beta$. Taking into account the
unitarity of the scattering matrix ${\bf s}$ we find, 
\begin{equation}
\nu (\alpha , {\bf r} ) = -\frac{1}{2\pi i}
\sum_{\beta} Tr \left[ {\bf s}_{\alpha \beta }^{\dagger}
\frac{\delta {\bf s}_{\alpha \beta }}{e\delta U({\bf r})} \right]\;\;.
\label{emit1}
\end{equation}
At the top of this pyramid of density of states is the local 
density of states. For the local density of states 
we impose no restriction as to where the carriers come from 
nor as to where they will go in the future. The local density
of states is thus the sum over all local partial density of states.
In particular it is also the sum of all injectivities or 
the sum of all emissivities,  
\begin{equation}
\nu ({\bf r} ) = \sum_{\beta} \nu ({\bf r},  \beta ) 
= \sum_{\alpha} \nu (\alpha , {\bf r} ) .
\label{ldos}
\end{equation}
and is thus given by 
\begin{equation}
\nu ({\bf r} ) = 
-\frac{1}{2\pi i}
\sum_{\alpha\beta} Tr \left[ s_{\alpha \beta }^{\dagger}
\frac{\delta s_{\alpha \beta }}{e\delta U({\bf r})} \right]\;\;.
\label{ldos1}
\end{equation}
We will later show the usefulness of these density of states. 
Here we only briefly mention the magnetic field symmetry.
Since in a uniform field the scattering matrix 
has the reciprocity symmetry ${\bf s}_{\alpha \beta }(B) = 
{\bf s}_{\alpha \beta }(-B)$,
we see that the partial density of states  have the symmetry
$\nu_{B} (\alpha , {\bf r},  \beta ) = 
\nu_{-B} (\beta , {\bf r}, \alpha )$.
Similarly, the injectivities and emissivities are 
related by reciprocity $\nu_{B} ({\bf r}, \alpha) = 
\nu_{-B} (\alpha, {\bf r})$ whereas the local 
density of states is an even function of magnetic field
$\nu_{B} ({\bf r}) = \nu_{-B} ({\bf r})$. 
Fig. 2 shows the local density of states, the injectivity 
and the ratio of the injectivity to the local density of states
for the simple case of a delta function barrier in an otherwise perfect
ballistic one-channel wire
connecting contacts $1$ and $2$. For additional simple examples 
we refer the reader to Refs. \onlinecite{GRAM,GASP,ZHAO}. 
Partial density of states in disordered structures 
are evaluated numerically in Ref. \onlinecite{DEJE} 
\begin{figure}
\narrowtext
\epsfysize5cm
\epsffile{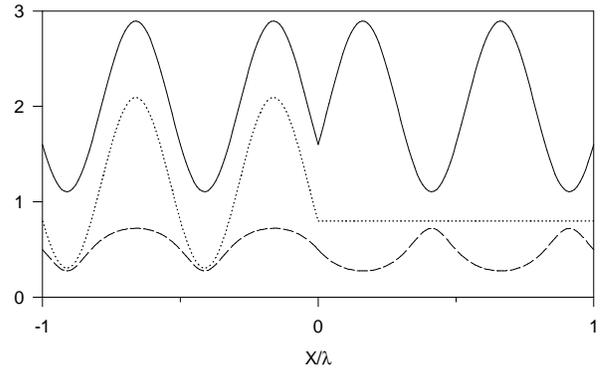}
\caption{The local density of states
$\nu(x)$ (solid line), the injectivity $\nu(x,1)$
of contact $1$ (dotted line) (both in units of $1/hv$), and the ratio of
these two densities
$\nu(x,1)/\nu(x)$ (dashed line) of a one dimensional wire with a $\delta$
barrier at $x=0$ which a strength that leads to a transmission 
probability of $T = 0.8$. The coordinate $x$ is in units of the Fermi 
wavelength $\lambda$. After Ref. \protect\onlinecite{GRAM}.} 
\label{oscfig}
\end{figure}
\section{Global partial density of states}

Often we are interested not in the density of 
states at a given point but in a certain volume
$\Omega$. Of course the partial density of states in 
$\Omega$ are obtained simply by integrating 
the local partial density of states over the 
volume of interest. Of particular interest 
are the density of states of the entire sample. 
In this case the surface of the volume $\Omega$
will intersect the leads connected to the sample. 
Suppose now that these intersections coincide 
with our specification of the scattering matrix. 
The phases which occur in the scattering matrix ${\bf s}_{\alpha\beta}$ 
are the phases which are accumulated through scattering 
between the intersection $S_{\alpha}$ of $\Omega$
in lead $\alpha$ and the intersection of $S_{\beta}$
in lead $\beta$. Under these circumstances 
a simplification of the above expressions 
is possible, since the functional 
derivative with respect to the local 
potential and the integration over $\Omega$
can be replaced by an energy derivative, 
\begin{equation}
- \int_{\Omega} d{\bf r}^{3} 
\frac{\delta }{e\delta U({\bf r})}
\rightarrow \frac{d}{dE} .
\label{wkb}
\end{equation}
For the {\it global} density of states (density of states
of the entire sample) we find the partial density of states, 
\begin{equation}
N (\alpha , {\bf r},  \beta ) = \frac{1}{4\pi i}
Tr \left[ {\bf s}_{\alpha \beta }^{\dagger}
\frac{d {\bf s}_{\alpha \beta }}{dE} - 
\frac{d {\bf s}_{\alpha \beta }^{\dagger}}{dE}
{\bf s}_{\alpha \beta }\right]\;\;
\label{gpdos2}
\end{equation}
from which we obtain the injectance of contact $\alpha$ 
\begin{equation}
N ({\bf r}, \beta ) = \frac{1}{2\pi i} 
\sum_{\alpha} Tr \left[ {\bf s}_{\alpha \beta }^{\dagger}
\frac{d {\bf s}_{\alpha \beta }}{dE} \right]\;\;
\label{inject3}
\end{equation}
the emittance into contact $\alpha$,
\begin{equation}
N(\alpha ) = \frac{1}{2\pi i}
\sum_{\beta} Tr \left[ {\bf s}_{\alpha \beta }^{\dagger}
\frac{d {\bf s}_{\alpha \beta }}{dE} \right]\;\;
\label{emit3}
\end{equation}
and the global density of states 
\begin{equation}
N =\sum_{\alpha} N(\alpha ) = \frac{1}{2\pi i}
\sum_{\alpha\beta} Tr \left[ {\bf s}_{\alpha \beta }^{\dagger}
\frac{d {\bf s}_{\alpha \beta }}{dE} \right] . 
\label{gdos}
\end{equation}

The scattering matrix is unitary and all its eigenvalues 
$\lambda_{i} = exp(i\zeta_{i})$
are thus on the unit circle.
In terms of these eigenvalues we have 
\begin{equation}
N = \frac{1}{2\pi} \sum_{i} \frac{d\zeta_{i}}{dE} = \frac{1}{2\pi}
\frac{d}{dE} \log[det({\bf s})]
\label{logdet}
\end{equation} 
where $det({\bf s})$ is the determinant of the scattering matrix. 

Eq. (\ref{logdet}) is a familiar expression of the 
density of states \cite{SMITH,DASH,STAF}. 
We recall it here to emphasize that it is an approximate expression 
for the density of states of a mesoscopic conductor. 
In nuclear scattering theory \cite{SMITH,DASH} it is 
turned into an exact expression by considering not the total 
density of states like we have done but the excess density of states
in comparison to a free problem
and by taking the volume of interest 
to infinity (The problem with and without 
a scattering object are compared). 
In contrast for our type of scattering problems 
(for instance for a multiprobe mesoscopic conductor) 
there is no reasonably defined free problem and 
similarly it would be meaningless to consider the volume of 
interest to become infinitely large. Therefore 
in mesoscopic physics 
the exact expression for the density of states of the 
sample is \cite{BTP2,GASP}  
\begin{equation}
N = \int_{\Omega} d{\bf r}^{3} \nu ({\bf r} ) = 
-\frac{1}{2\pi i} \int_{\Omega} d{\bf r}^{3} 
\sum_{\alpha\beta} Tr \left[ s_{\alpha \beta }^{\dagger}
\frac{\delta s_{\alpha \beta }}{e\delta U({\bf r})} \right]\;\;
\label{Nmeso}
\end{equation}
whereas the more familiar expression Eq. (\ref{logdet})
is an approximation. 

Except in very special cases 
the two expressions for the local density 
need not to be distinguished. However the difference is 
important when WKB 
does not hold. For instance using Eq. (\ref{logdet}),  
the density of states 
inside a rectangular tunneling barrier of height $U_{0}$
diverges like $E^{-1/2}$ as the energy kinetic energy 
$E$ of the incident particles tends to zero whereas 
if one uses Eq. (\ref{Nmeso}) one finds (the much more 
reasonable result) that the density of states is finite 
and vanishes like $E^{1/2}$ for small energies \cite{LARMOR}.  
The distinction between energy derivatives and potential 
derivatives is familiar from the discussion of 
characteristic tunneling times \cite{BL82,LARMOR,LEAE1,LEAE2}. 
For a discussion of the 
connection between the local partial density of states 
and characteristic tunneling times we refer the reader 
to Ref. \onlinecite{MB01}.

\section{Absorption and Emission of Particles: Injectivities
and Emissivities}

We next present a simple example which illustrates 
the usefulness of the densities of states introduced above. 
Suppose that the potential has 
in some small volume element $\Omega$ a small imaginary 
part \cite{MB90,KUMAR,CWJB}. 
Thus the potential is $U({\bf r})$ out-side $\Omega$
but is 
$U({\bf r})-i\hbar \Gamma$ inside $\Omega$. 
To proceed we take into account that the scattering matrix 
is a functional of the potential and expand the scattering 
matrix to first order in $\Gamma$. 
This gives for the 
transmission and reflection probabilities in the presence of 
a small absorption \cite{MB90}                     
\begin{equation}
T_{\alpha \beta}(\Gamma ) = T_{\alpha\beta}
\left(1- \int_{\Omega}\; d{\bf r}\; \Gamma \;\nu(\alpha,{\bf r},\beta) \right)
\label{g3}
\end{equation}
where $T_{\alpha\beta}$
is the transmission probability of the scattering problem 
without absorption if $\alpha$ and $\beta$ are different.
If $\alpha$ and $\beta$ are equal, we find a reflection probability  
\begin{equation}
R_{\alpha \alpha}(\Gamma ) = R_{\alpha\alpha}
\left(1- \int_{\Omega} d{\bf r}\; 
\Gamma  \; \nu(\alpha,{\bf r},\alpha )  \right) .
\label{g4}
\end{equation}
Thus in the presence of absorption 
the reflection and transmission probabilities 
are "diminished" according to the partial density of states.  
Consider now the carriers streaming from 
contact $\beta$ into the sample in a narrow 
energy interval $dE$.
The incident current
$I_{in}$ must be equal to the sum of
the transmitted current $I_{T}$, the reflected current $I_{R}$
and the absorbed current $I_{\Gamma}$, 
\begin{equation} 
I_{in} = I_{T} + I_{R} + I_{\Gamma} .
\label{g5}
\end{equation}
Taking into account that the total incident current from contact 
$\alpha$ is $I_{\beta} = (e/h) M_{\beta} dE$, 
and taking into account that $\sum_{\alpha \ne \beta} T_{\alpha\beta}
+ R_{\beta\beta} = M_{\beta}$ we find that the total abosrbed flux is 
given by the sum  
\begin{equation} 
I_{\Gamma} =\int_{\Omega} d{\bf r} \; \Gamma  \;\nu({\bf r},\beta).
\label{g6}
\end{equation}
Thus the absorbed flux is determined by the injectivity 
of contact $\beta$ into the volume of interest. 

Another way of determining the absorbed flux proceeds as follows. 
The absorbed flux is proportional to the integrated density 
of particles in the region of absorption.
The density of particles can be found from the scattering 
state $\psi_{\beta m}({\bf r})$ which describes carriers incident
from contact ${\beta}$ in channel ${ m}$. The wave function is 
normalized such that it has unit amplitude in the incident channel. 
We assume that all incident channels in contact $\beta$ are filled
equally. 
The absorbed flux is thus \cite{MB90} 
\begin{equation} 
I_{\Gamma}(\beta)  = \int d{\bf r} \; \Gamma \;\sum_{m} 
\frac{1}{hv_{\beta}}|\psi_{\beta m}({\bf r})|^{2} .
\label{g7}
\end{equation}
Note that here the density of states $1/hv_{\beta m}$ of the 
asymptotic scattering region appears. $v_{\beta m}$ is 
the velocity of carriers in the incident channel. 
Thus we have found a wave function representation for the 
injectivity. Comparing Eq. (\ref{g4}) and Eq. (\ref{g3}) gives
\begin{equation} 
\nu({\bf r},\beta) = 
\sum_{m} \frac{1}{hv_{\beta m}}|\psi_{\beta m}({\bf r})|^{2} .
\label{g8}
\end{equation}
The total local density of states $\nu({\bf r})$ at point ${\bf r}$ is 
obtained by considering carriers incident from both contacts. 

There is now an interesting additional problem to be addressed. 
Instead of a potential which acts as a carrier sink (as an absorber)
we can ask about a potential which acts as a carrier source. Obviously, 
all we have to do to turn our potential into a carrier source is to 
change the sign of the imaginary part of the potential. As a consequence 
of a carrier source in the volume $\Omega$ we should observe
a particle current into all contacts of our sample.  
To find these currents we consider a small energy interval at the 
energy of interest.                   
The total current injected into the sample in $\Omega$ is 
\begin{equation} 
I_{in} (y) = \int_{\Omega} d{\bf r} \;  \Gamma  \; \nu({\bf r}). 
\label{g11}
\end{equation}
Taking into account that the incident current is normalized to $1$
the current $I_{out}(\beta)$  in contact $\beta$ due to a carrier 
source in $\Omega$ is given by 
\begin{equation}
I_{out}(\alpha) = (e/h) dE 
[M_{\alpha}- R_{\alpha\alpha} (\Gamma)
- \sum_{\beta \ne \alpha} T_{\alpha\beta}(\Gamma)]
\label{g12}
\end{equation}
due to the modification of both the transmission and reflection coefficients.
Using Eqs. (\ref{g3}) and (\ref{g4})  
(with $\Gamma$ replaced by -$\Gamma$)
gives 
\begin{equation}
I_{out}(\alpha) = - \int_{\Omega} d{\bf r}\; \Gamma \;\sum_{\beta}
\nu(\alpha,{\bf r},\beta) 
= - \int_{\Omega}  d{\bf r}\; \Gamma \;\nu(\alpha,{\bf r}) . 
\label{g13}
\end{equation}
The current in contact $\alpha$ is determined by the 
{\it emissivity} $\nu(\alpha,{\bf r})$ of the region
$\Omega$ into contact $\alpha$.

We conclude this section with a remark of caution.
The partial densities are not necessarily positive
as we would expect from a usual density of states. 
(The simple example of an (asymmetric) resonant double barrier
shows that  
one of the diagonal elements $\nu(\alpha,{\bf r},\alpha)$
has a range of energies where it is negative \cite{MB90}. 
Note that this implies that a small absorption inside a 
sample can {\it increase} a reflection probability).
The fact that the partial density of states are not necessarily 
positive is a consequence of over specification: both the incident 
and the exiting channel are prescribed. 
In contrast, the injectivities and emissivities are, however, always positive. 
The proof is given by Eq. (\ref{g6}).

\section{Generalized Bardeen formulae}

It is well known that with a scanning tunneling microscope (STM)
we can measure the local density of states \cite{STM}. STM measurements
are typically performed in a two terminal geometry, in which
the tip of the microscope represents one contact and the sample 
provides another contact \cite{STM}. Here we are interested 
in the transmission 
probability from an STM tip into the contact of a sample with two or more 
contacts as shown in Fig. \ref{eintip}. Geometries which use 
more than one tunneling tip are also of interest \cite{STM2,STM3}. 
Thus we deal with a multiterminal 
transmission problem \cite{GRAM}. 
If we denote the contacts of the sample 
by a Greek letter $\alpha = 1, 2, ..$ and use {\it tip} 
to label the contact of the STM tip, we are interested in the 
tunneling probabilities $T_{\alpha tip}$ from the tip 
into contact $\alpha$ of the sample. In this case the 
STM tip acts as carrier source. Similarly we ask about 
the transmission probability $T_{tip \alpha}$ 
from a sample contact to the tip. In this case the STM tip 
acts as a carrier sink. Earlier work has addressed this problem 
either with the help of electron wave dividers \cite{ENGQ,MB89}, 
or by applying the Fermi Golden Rule \cite{IMRY}. 
Recently, Gramespacher and 
the author \cite{GRAM} have returned to this problem and have derived 
expressions for these transmission probabilities from the
scattering matrix of the full multiprobe problem \cite{MB86} (sample plus tip). 
For a tunneling contact with a density of states $\nu_{tip}$
which couples locally at the point 
$x$ with a coupling energy 
$|t|$ these authors found
\begin{equation}
T_{tip,\alpha}=4\pi^2\nu_{tip}|t|^2\nu(x,\alpha)\, ,\label{w1}
\end{equation}
\begin{equation}
T_{\alpha tip}=4\pi^2\nu(\alpha,x)|t|^2\nu_{tip}\, .\label{w2}
\end{equation}
In a multiterminal sample the transmission probability
from a contact $\alpha$
to the STM tip is given by the injectivity of contact $\alpha$ into the 
point $x$ and the transmission probability from the tip 
to the contact $\alpha$ is given by the emissivity of the point $x$
into contact $\alpha$. Eqs. (\ref{w1}) and (\ref{w2})
when multiplied by the unit of conductance $e^{2}/h$ 
are generalized Bardeen conductances for multiprobe 
conductors. Since the local density of states of the tip
is an even function of magnetic field 
and since the injectivity and emissivity are related by reciprocity
we also have the reciprocity relation $T_{tip,\alpha}(B) = 
T_{\alpha,tip} (-B)$. 

The presence of the tip also affects transmission and reflection 
at the massive contacts of the sample. 
To first order in the coupling energy $|t|^{2}$
these probabilities are 
given by 
\begin{equation}
T^{tip}_{\alpha\beta}= T_{\alpha\beta} - 
4\pi^2\nu(\alpha,x,\beta)|t|^2\nu_{tip}\, .\label{w3}
\end{equation}
Here the upper index $tip$ indicates that 
this transmission probability 
is calculated in the presence of the 
tip. $T_{\alpha\beta}$ without an upper index
is the transmission probability in the absence of the tip. 
The correction to the transmission probabilities $\alpha \ne \beta$
and reflection probabilities $\alpha = \beta$ is determined by 
the partial densities of states, the coupling energy and the 
density of states in the tip. Note that if these probabilities
are placed in a matrix then each row and each column of this 
matrix adds up to the number of quantum channels in the contacts.

\section{Voltage probe and inelastic scattering}

Consider a two probe conductor much smaller than any inelastic or phase
breaking length. The conductance of such a conductor can then be said 
to be coherent and its conductance is at zero
temperature given by the Landauer formula 
$G = (e^{2}/h) T$, where $T$ is the 
probability for transmission form one contact to the other. 
How is this result affected by events which break the phase or by 
events which cause inelastic scattering? To investigate this 
question Ref. \onlinecite{MB88} proposes to use 
an additional (third) contact to the sample. 
The third probe acts as a voltage probe which has its potential 
adjusted in such a way that there is no net current  
flowing into this additional probe,
$I_{3} = 0$. The current at the third probe is set to zero by floating
the voltage $\mu_3 = eV_{3}$ at this contact to a 
value for which $I_{3}$ vanishes. 
The third probe acts, therefore, like a {\it voltage probe}.
Even though the total current at the voltage probe vanishes individual 
carriers can enter this probe if they are at the same time replaced 
by carriers emanating from the probe \cite{MB88}. 
Entering and leaving a contact are 
{\it irreversible} processes, since there is no definite phase relationship
between a carrier that enters the contact and a carrier that leaves 
a contact. In a three probe conductor, the relationship 
between currents and voltages is given by $I_{\alpha} = 
\sum_{\beta} G_{\alpha\beta}V_{\beta}$ where the $G_{\alpha\beta}$
are the conductance coefficients. Using the condition $I_{3} =0$
to find the potential $V_{3}$ and eliminating this potential
in the equation for $I_{2}$ or $I_{1}$ gives for the two probe conductance
in presence of the voltage probe 

\begin{equation}
G = - (G_{21} + \frac{G_{23}G_{31}}{G_{31}+G_{32}}) .
\label{v1}
\end{equation}

For a very weakly coupled voltage probe (see Fig. \ref{eintip}) 
we can use Eqs. (\ref{w1} - \ref{w3}).
Taking into account that $G_{\alpha\beta} = - (e^{2}/h) 
T^{tip}_{\alpha\beta}$ 
for $\alpha \ne \beta$
we find 
\begin{equation}
G = \frac{e^{2}}{h} (T - 4\pi^2 |t|^2 [\nu(2,x,1) - 
\frac{\nu(2,x)\nu(x,1)}{\nu(x)}] ) . 
\label{v2}
\end{equation} 
Here $\nu(x)$ is the local density of states at the location of the
point at which the voltage probe couples to the conductor. 
Eq. (\ref{v2}) has a simple interpretation \cite{MB88}. 
The first term $T$ is 
the transmission probability of the conductor in the absence of the 
voltage probe. The first term inside the brackets 
proportional to the local partial density of states gives the 
reduction of coherent transmission due to the presence of the 
voltage probe. The second term in the brackets is the incoherent
contribution to transport due to inelastic scattering induced by the 
voltage probe. It is proportional to the injectivity of contact $1$
at point $x$. A fraction $\nu(2,x)/{\nu(x)}$ of the carriers which reach
this point, proportional to its emissivity, are scattered forward 
and, therefore, contribute to transport. Notice the different signs
of these two contributions. The effect of inelastic scattering (or dephasing)
can either enhance transport or diminish transport, depending on 
whether the reduction of coherent transmission (first term) 
or the increase due to incoherent transmission (second term) 
dominates.

Instead of a voltage probe, we can also use 
an optical potential to simulate inelastic scattering or dephasing. 
However, in order to preserve current, we must use both an absorbing 
optical potential (to take carriers out) and an emitting optical 
potential (to reinsert carriers). The absorbed and re-emitted 
current must again exactly balance each other \cite{PBCB}. 
From Eq. (\ref{g4}) it is seen that the coherent current is again 
diminished by $\Gamma \nu(2,x,1)$, i. e. by the partial density 
of states at point $x$. The total absorbed current is proportional 
to $\Gamma \nu(x,1)$, the injectance of contact $1$ into this point.
As shown in section 3 a carrier emitting optical potential 
at $x$ generates a current $ - \Gamma \nu(1,x)$ in contact $1$
and generates a current $ - \Gamma \nu(2,x) $ in contact $2$.
It produces thus a total current $ - \Gamma \nu(x) $.
In order that the generated and the absorbed current are equal 
we have to normalize the emitting optical potential \cite{MB01}  
such that it generates a total current proportional to
$\Gamma \nu(x,1)$. The current produced by an optical potential
normalized in such a way is thus $-\Gamma \nu(2,x)\nu(x,1)/\nu(x)$. 
The sum of the two contributions \cite{MB01}, the absorbed current and the 
re-emitted current gives an overall transmission (or conductance)
which is given by Eq. (\ref{v2}) with $4\pi |t|^{2}$ replaced by 
$\Gamma$.   
 
Thus the weakly coupled voltage probe 
(which has current conservation built in)
and a discussion based on optical potentials coupled with a 
current conserving re-insertion of carriers are equivalent \cite{PBCB}.
There are discussions in the literature which invoke optical
potentials but do not re-insert carriers. Obviously, such 
discussions violate current conservation. 
Even if current is conserved, the partitioning of the reinserted carriers
has to be done as described above. Otherwise such fundamental 
properties like the Onsager symmetry are violated \cite{JAY}. 
 
We would like to point out that there is in general no reason that 
tunnel contacts and absorption and re-emission with the help
of imaginary potentials lead to identical results. The equivalence     
rests on a particular description of the voltage probe.  There are 
many different ways of coupling a tunnel 
contact  and our description
of the voltage probe given here is not unique. The claim can only be that 
for sufficiently weak optical absorption and re-insertion 
of carriers there exits {\it one} voltage probe model 
which gives the same answer.

\section{The Friedel sum rule} 

The connection between the phase shift of scattering 
elements and the charge is known as Friedel sum rule. 
It is most often stated as a global relation 
between the charge in an energy interval $dE$ 
and the derivatives of the scattering phases with respect to 
energy as 
\begin{equation}
dQ/dE = \frac{e}{2\pi} \sum_{i} \frac{d\zeta_{i}}{dE} = 
\frac{e}{2\pi} \frac{d}{dE} \log[det({\bf s})] .
\label{friedel1}
\end{equation} 
From the discussion at the end of Section II it is evident
that this relation is not exact since in the applications 
envisioned here the scattering matrix and the phases
are calculated for a finite volume $\Omega$. Thus whenever 
WKB does not hold Eq. (\ref{friedel1}) will not hold. 
The exact relation follows from Eq. (\ref{g8}) and Eqs. (\ref{ldos})
and (\ref{ldos1}) and is given by 
$$
\nu({\bf r}) = 
\sum_{\beta m} \frac{1}{hv_{\beta m}}|\psi_{\beta m}({\bf r})|^{2} 
= -\frac{1}{2\pi i} 
\sum_{i} \frac{\delta \zeta_{i}}{e\delta U({\bf r})}
$$
\begin{equation} 
\label{friedel2}
= -\frac{1}{2\pi i}
\sum_{\alpha\beta} Tr \left[ s_{\alpha \beta }^{\dagger}
\frac{\delta s_{\alpha \beta }}{e\delta U({\bf r})} \right]\; .
\end{equation}
This relation is exact {\it locally} and by integration 
can be applied to any volume of interest. Application 
of Eq. (\ref{friedel1}) to situations 
in which WKB does not hold has already given rise to statements 
in the literature that the Friedel sum rule does not apply in 
mesoscopic conductors \cite{DEO}.

\section{The charge response to an internal potential} 

We next discuss problems 
for which the self-consistent adjustment of the potential 
is important. With this goal in mind we first investigate 
the variation of the local carrier density $\delta n({\bf r})$
in response to a local potential variation $\delta U({\bf r}^{\prime})$ 
at a point ${\bf r}^{\prime}$. 
This variation is given by the Lindhard function, 
$\delta n({\bf r}) = - \int d{{\bf r}^{\prime}}^{3} 
\Pi({\bf r}, {\bf r}^{\prime}) e\delta U({\bf r}^{\prime})$.
This function plays an important role if we want to determine 
the carrier distribution self-consistently. Since the 
local charge density at 
equilibrium is $n({\bf r}) = \int dE \nu({\bf r}) f(E)$,
it follows immediately that the Lindhard function 
is given by 
\begin{equation}
\Pi({\bf r},{\bf r}^{\prime}) = - \int dE \frac{\delta \nu({\bf r})} 
{e\delta U({\bf r}^{\prime})}f(E) .
\label{lind1}
\end{equation}
This response now involves second order 
functional derivatives of the scattering matrix. 
It can be simplified, if we are interested in potential 
variations not on the microscopic scale but only over 
distances large to a Fermi wave length $\lambda_F$.
In that case we can integrate the Lindhard function  
over its second argument over a volume that is 
large compared to a Fermi wave length
and use the WKB formula which replaces the functional 
derivative with respect to $ U({\bf r}^{\prime})$
with an energy derivative. This gives 
\begin{equation}
\Pi({\bf r},{\bf r}^{\prime}) = 
\delta({\bf r}- {\bf r}^{\prime}) \int dE \frac{d \nu({\bf r})} 
{dE}f(E) .
\label{lind2}
\end{equation}
Partial integration with respect to the energy $E$ gives 
\begin{equation}
\Pi({\bf r}^{\prime},{\bf r}) = 
 \delta({\bf r}^{\prime} - {\bf r}) \int dE \nu({\bf r})
\left(-\frac{df(E)}{dE}\right) .
\label{lind3}
\end{equation}
Over sufficiently large distances the Lindhard function 
is determined by the local density of states.

\section{The local charge response of a mesoscopic conductor} 

Consider a small deviation $\delta \mu_{\alpha} = \mu_{\alpha} - \mu$
of the electrochemical potential 
$\mu_{\alpha}$ of contact ${\alpha}$ away from its equilibrium 
value $\mu$. At fixed internal potential this increases 
the carrier density inside the conductor by       
$\delta n({\bf r}) = \nu({\bf r},\alpha) \delta\mu_{\alpha}$.
This injected charge generates 
due to the Coulomb interaction an electrostatic response 
$\delta U({\bf r})$ which via the Lindhard function 
will also give a contribution to the charge density. 
Thus the local carrier density is \cite{MB93}  
\begin{equation}
\delta n({\bf r}) = \sum_{\alpha} \nu({\bf r}, \alpha)\delta\mu_{\alpha}
- \int d{\bf r^{\prime}}^{3}
\Pi({\bf r}^{\prime},{\bf r}) \delta U({\bf r}^{\prime}).
\label{den1}
\end{equation}
Suppose that all potentials $\delta\mu_{\alpha}$ applied 
to the conductor are rised by the same amount 
$\delta\mu$. Since the state of a conductor 
can only be changed by voltage differences 
it follows that the electrostatic potential 
increases also by $\delta U({\bf r}^{\prime}) = \delta\mu$
and that $\delta n({\bf r}) = 0$. As a consequence 
of this gauge invariance we have \cite{MB93}
\begin{equation}
\sum_{\alpha} \nu({\bf r}, \alpha) = \nu({\bf r})
= \int d{\bf r^{\prime}}^{3}
\Pi({\bf r}^{\prime},{\bf r}).
\label{gauge}
\end{equation}
The spatial integral over one of the arguments of the 
Lindhard function is equal to the local density of states. 
Note that Eq. (\ref{lind1}) is a symmetric function of its 
spatial arguments. 

To bring the determination of the local charge response 
to an end, we must now insert Eq. (\ref{den1})
into the Poisson equation and solve for the potential 
$\delta U({\bf r}^{\prime})$. 
Suppose that a test charge at point ${\bf r}^{\prime}$
generates at point ${\bf r}$ a potential 
$g({\bf r},{\bf r}^{\prime})$. Here $g({\bf r},{\bf r}^{\prime})$
is the Green's function of the Poisson equation with a
non-local screening kernel determined by the Lindhard function 
mentioned above. The injected charge $e \nu({\bf r}, \alpha)\delta\mu_{\alpha}$
generates the potential $u_{\alpha}({\bf r}) \delta\mu_{\alpha}$
where the characteristic potential \cite{MB93} $u_{\alpha}({\bf r})$
is given by
\begin{equation}
u_{\alpha}({\bf r}) = 
\int d{\bf r^{\prime}}^{3} g({\bf r},{\bf r}^{\prime})
\nu({\bf r}^{\prime}, \alpha).
\label{char}
\end{equation}
The injectivity is the source term of the characteristic potential 
$u_{\alpha}({\bf r})$. 
With the help of the characteristic potentials the internal 
potential can be expressed in linear response to the applied 
electrochemical potentials. Using this, 
the local carrier distribution 
is found to be 
\begin{equation}
\delta n({\bf r}) =
\sum_{\alpha} \nu^{tot}({\bf r}, \alpha) \delta\mu_{\alpha}
\label{den3}
\end{equation}  
with a screened density of states \cite{MB93} 
\begin{equation}
\nu^{tot}({\bf r}, \alpha) = 
\nu({\bf r}, \alpha)
- \int d{\bf r^{\prime}}^{3} d{\bf r^{\prime\prime}}^{3}
\Pi({\bf r},{\bf r}^{\prime}) g({\bf r}^{\prime},{\bf r}^{\prime\prime})
\nu({\bf r^{\prime\prime}}, \alpha).
\label{den4}
\end{equation} 
Eq. (\ref{den3}) provides a gauge invariant description of the 
local carrier density in response to small variations of the 
potentials applied to the contacts of the sample. 
The total variation of the charge is zero, 
$\int_{\Omega} d{\bf r} \delta n({\bf r}) = 0$, 
for a volume $\Omega$ that is sufficiently large (encloses
the entire sample). 

\section{Dynamic Conductance} 

In this section we discuss as an additional application 
of partial density of states briefly the ac-conductance of 
mesoscopic systems. We consider a conductor with 
an arbitrary number of 
contacts labeled 
by a Greek index $\alpha = 1, 2, 3...$. 
The problem is to find the relationship 
between the currents $I_{\alpha}(\omega)$ at frequnecy $\omega$
measured at the contacts of the sample 
in response to a sinusoidal voltage with amplitude $V_{\beta}(\omega)$
applied to contact $\beta$. The relationship between currents 
and voltages is given by a dynamical conductance matrix \cite{BTP2,MB93}
$G_{\alpha\beta} (\omega)$ such that 
$I_{\alpha}(\omega) = \sum_{\beta}G_{\alpha\beta} (\omega)V_{\beta}(\omega)$.
All electric fields are localized in space. 
The overall charge on the conductor is conserved. 
Consequently, current is also conserved and the currents 
depend only on voltage differences. Current conservation
implies  $\sum_{\alpha}G_{\alpha\beta} = 0$ for each $\beta$.  
In order that only voltage differences matter, the dynamical
conductance matrix has to obey $\sum_{\beta}G_{\alpha\beta} = 0$
for each $\alpha$. We are interested here in the low frequency 
behavior of the conductance and therefore we can expand the conductance in
powers of the frequency \cite{MB93},  
\begin{equation}
G_{\alpha\beta} (\omega) = G^{0}_{\alpha\beta} - i \omega E_{\alpha\beta} 
+ K_{\alpha\beta} \omega^{2} + O(\omega^{3}) .
\label{ac1}
\end{equation} 
Here $G^{0}_{\alpha\beta}$ is the dc-conductance matrix. 
$E_{\alpha\beta}$ is called the {\it emittance} matrix 
and governs the displacement currents. $K_{\alpha\beta}$ 
gives the response to second order in the frequency. 
All matrices $G^{0}_{\alpha\beta}, E_{\alpha\beta}$ and 
$K_{\alpha\beta}$ are real. 

We focus here on the emittance matrix $E_{\alpha\beta}$.
The conservation of the total charge can only be achieved 
by considering the long-range Coulomb interaction. 
To first order in frequency it is precisely the charge 
distribution given by Eq. (\ref{den3}) 
which counts. Ref. \onlinecite{MB93} finds for the 
emittance matrix 
\begin{equation}
E_{\alpha\beta} = e^{2} [ \int d{\bf r} \nu(\alpha,{\bf r},\beta) - 
\int d{\bf r}^{\prime} d{\bf r}
\nu(\alpha,{\bf r}^{\prime})g({\bf r}^{\prime},{\bf r}) \nu({\bf r},\beta) ] 
\label{ac2}
\end{equation} 
Here the first term proportional to the integrated 
partial density of states is the ac-response at low frequencies
which we would have in the absence of interactions. 
The second term has the following simple interpretation:
an ac-voltage applied to contact $\beta$ would (in the absence of
interactions) lead to a charge built up at point ${\bf r}$ given by 
the injectivity of contact $\beta$. Due to interaction, this charge generates 
at point ${\bf r}^{\prime}$ a variation in the local potential which 
then induces 
a current in contact $\alpha$ proportional to the emissivity 
of this point into contact $\alpha$. 
Note, that if screening is local (over a length scale 
of a Thomas Fermi wave length) we have $g({\bf r}^{\prime},{\bf r})= 
\delta ({\bf r}^{\prime} - {\bf r}) \nu^{-1}({\bf r})$. 
In this limit the close connection between Eq. (\ref{ac2}) 
and Eq. (\ref{w2}) is then obvious. We have to refer the 
reader to a review \cite{MBTC} for a more extended discussion and 
for references to specific applications of Eq. (\ref{ac2}).

A closely related development which could be discussed here is a theory of 
quantum pumping in small systems. 
In quantum pumping one is interested in the current generated as 
two parameters (like gate voltages, magnetic flux) 
which modulate the system are varyed sinusoidally but out of phase. 
Brouwer \cite{BR1}, Avron et al. \cite{AVRON}, and 
Polianski and Brouwer \cite{BR2} develop a theory which 
is based on the slow modulation of the partial densities of states.

\section{Partial Density of States Matrix}

Thus far we have been interested in the charge distribution 
inside a mesoscopic conductor which is established in response
to changes in external parameters, such as a voltage applied 
to the contact of the sample. Similar to the dynamic fluctuations 
of the current away from its average value, we can also 
ask about the fluctuations of the charge away from its average 
value. Spontaneous dynamic fluctuations of current and 
charge can have many sources in an electrical conductor. 
Here we are concerned with contributions 
which are fundamental in the sense that they can not be avoided. 
Nyquist or thermal noise can not be avoided at any elevated 
temperature. A second source, shot noise
results from the granularity 
of charge (the charge quantization) and the quantum mechanical 
uncertainty which arises whenever there are different possible 
outcomes of a scattering process\cite{BB} (such as the possibility of 
a particle to be either reflected or transmitted).

To describe the charge fluctuations due to carriers 
in the conductor \cite{MATH}, we consider the 
Fermi-field 
\begin{equation}
\label{fermi}
\hat{\Psi} ({\bf r},t)  = \sum_{\alpha m} \int dE 
\frac{\psi_{\alpha m} ({\bf r},E)}{(hv_{\alpha m} (E))^{1/2}} 
\hat{a}_{\alpha m} (E) exp(-iEt/\hbar), 
\end{equation}
which annihilates an electron at point ${\bf r}$ and time $t$.
The Fermi field Eq. (\ref{fermi}) is built up 
from all scattering states $\psi_{\alpha m} ({\bf r},E)$
which have unit incident amplitude in contact $\alpha$ in channel
$m$. The operator $\hat{a}_{\alpha m} (E)$ annihilates 
an incident carrier in reservoir $\alpha$ in channel
$m$. $v_{\alpha m}$ is the velocity in the incident channel $m$ in 
reservoir $\alpha$. 
The local carrier density at point ${\bf r}$ and time $t$
is determined by $\hat{n}({\bf r},t) =  \hat{\Psi}^{\dagger}({\bf r},t)
\hat{\Psi}({\bf r},t)$. 
We will investigate 
the density operator in the frequency domain, $\hat{n}({\bf r},\omega)$.
Using the Fermi-field we find, 
$$
\hat{n}({\bf r},\omega) =  
\sum_{\alpha m \beta n} \int dE 
\frac{1}{(hv_{\alpha m} (E))^{1/2}}
\frac{1}{(hv_{\beta n} (E + \hbar \omega))^{1/2}}
$$
\begin{equation}\label{denp1}
\psi_{\alpha m}^{\ast} ({\bf r},E)  
\psi_{\beta n} ({\bf r},E + \hbar \omega) 
\hat{a}_{\alpha m}^{\dagger} (E) 
\hat{a}_{\beta n}(E + \hbar \omega) .
\end{equation}
This equation defines a {\it density of states matrix} 
${\bf \nu}_{\beta \gamma} ({\bf r})$
with elements 
\begin{eqnarray}\label{denp2}
{\nu}_{\gamma \delta  mn}({\bf r}, E, E + \hbar \omega) = && h^{-1}
(v_{\gamma m} (E) v_{\delta n} (E + \hbar \omega))^{-1/2} \nonumber \\
&& \psi_{\gamma m}^{\ast} ({\bf r},E) 
\psi_{\delta  n} ({\bf r},E + \hbar \omega) .
\end{eqnarray} 
If we form a vector ${\bf a}_{\alpha}$ with the annihilation opertors
of the (incoming) channels in contact ${\bf a}_{\alpha}$
the local density operator in the low frequency limit 
can be expressed with the help of this matrix 
in the following way,  
\begin{eqnarray}
\label{ldosm}
\hat{n}({\bf r},\omega) =
& &\sum_{\alpha \beta} \int dE 
\hat{\bf a}_{\alpha}^{\dagger} (E) {\bf \nu}_{\alpha \beta}({\bf r}) 
\hat{\bf a}_{\beta}(E + \hbar \omega) .
\end{eqnarray}
It is now very convenient and instructive to consider an expression 
for the density operator not in terms of wave functions 
but more directly in terms of the scattering matrix. 
In the zero-frequency limit, 
Ref. \onlinecite{MATH} introduces the matrix 
\begin{equation} 
{\bf \nu}_{\beta \gamma} (\alpha, {\bf r}) = 
- \frac{1}{4\pi i}
[{\bf s}^{\dagger}_{\alpha\beta}
\frac{\delta {\bf s}_{\alpha \gamma}}{e\delta U({\bf r})} -
\frac{\delta {\bf s}^{\dagger}_{\alpha \beta}}{e\delta U({\bf r})} 
{\bf s}_{\alpha \gamma}] .
\label{elmdef}
\end{equation} 
All quantities in this expression are evaluated at the Fermi energy $E$.
The matrix elements of Eq. (\ref{denp2}) 
are connected to the matrices ${\bf \nu}_{\beta \gamma} (\alpha, {\bf r})$ 
(and thus to wave functions) via \cite{MATH} 
\begin{equation}
{\bf \nu}_{\gamma \delta}({\bf r}) = \sum_{\alpha}
{\bf \nu}_{\beta \gamma} (\alpha, {\bf r})
= -\frac{1}{2\pi i} \sum_{\alpha}
\left( s_{\alpha \beta}^{\dagger}
\frac{\delta s_{\alpha \gamma}}{e\delta U({\bf r})} \right)\;.
\label{fdos3}
\end{equation}
Eq. (\ref{fdos3}) is 
a {\it local} Wigner-Smith life-time matrix \cite{SMITH}.

Eq. (\ref{fdos3}) was 
given in Ref. \onlinecite{MATH} and a derivation of this 
relation with the help of an optical potential is 
given in detail in Ref. \onlinecite{ANKA}. The relation between 
scattering states and the Green's function of the Schroedinger equation 
on one hand and its connection to the scattering matrix on the other hand 
can also 
be used \cite{LEVIN2,ENTIN}.
Recently Schomerus et al. \cite{SCHOM} proposed still another way 
to proceed by  using not the overall scattering matrix but scattering 
matrices for different portions of the system.

It is important to realize that in an electrical conductor 
charge fluctuations are not free but are screened like 
the average charge density. Inserting Eq. (\ref{ldosm})
into the Poisson equation with the non-local screening kernel 
determines a potential operator \cite{MATH}
\begin{eqnarray}
\label{pot1}
\hat{u}_{\alpha \beta}({\bf r},\omega) = \int d{\bf r}^{\prime} 
g({\bf r},{\bf r}^{\prime})  
\hat{\bf a}_{\alpha}^{\dagger} (E) {\bf \nu}_{\alpha \beta}({\bf r}^{\prime}) 
\hat{\bf a}_{\beta}(E + \hbar \omega) .
\end{eqnarray}
In random phase approximations the potential fluctuations 
give rise to induced charge fluctuations. Thus the true 
low frequency charge fluctuations in a conductor 
are determined by the contribution of the bare charge fluctuations 
and the induced charge fluctuations. This gives rise to 
a charge fluctuation 
operator \cite{MATH}
\begin{eqnarray}
\label{pot2}
\hat{\nu}^{tot}_{\alpha \beta}({\bf r}, \omega) = 
\hat{\nu}_{\alpha \beta}({\bf r}, \omega) - \int d{\bf r}^{\prime} 
\Pi({\bf r}, {\bf r}^{\prime})
\hat{{\bf u}}_{\alpha \beta}({\bf r}^{\prime},\omega) . 
\end{eqnarray}
Eq. (\ref{pot2}) permits a gauge invariant dsicussion of 
charge fluctuations in mesoscopic conductors. Eq. (\ref{pot2})
has been used in Ref. \onlinecite{MATH} to obtain the second order 
in frequency term of the ac-conductance (see Eq. (\ref{ac1})).
For a capacitive structure this defines a novel resistance
which determines the dissipative effects in charging and 
decharging a mesoscopic conductor. Pedersen and van Langen 
and the author \cite{PLB}
investigated the current induced into a nearby 
gate due to charge fluctuations in quantum point contacts and 
chaotic cavities. More recently, the charge fluctuations in 
two nearby mesoscopic conductors \cite{BUKS1} was treated 
and the effect of quantum dephasing due to charge fluctuations 
was calculated within this approach \cite{MBAM,ANKA}. 
Often self-consistent effects are neglected \cite{LEVIN,LEVIN2}. 
Charge correlations at two contacts capacitively coupled to 
a mesoscopic structure are investigated in Ref. \onlinecite{AMMB}.

Integration of Eq. (\ref{fdos3}) over the volume of the 
sample gives the global Wigner-Smith life-time matrix. 
In contrast to the local Wigner-Smith matrix the properties
of the (global) Wigner-Smith  life-time 
matrix have received wider attention. 
In recent years the focus has been on the calculation of the entire 
distribution function 
of delay times \cite{SOMMERS} for structures whose dynamics 
is in the classical limit chaotic (chaotic cavities)
and of one-dimensional wires \cite{OLDKU,TEXIER,SCHOM}. 
While in many cases an analysis of the global density 
is quite sufficient, in disordered structures, since screening is local, 
it is clearly desirable to proceed from the local 
density and to investigate the properties 
of the local Wigner-Smith matrix given here.

\section{Discussion}

We have  discussed a hierarchy of density of states as they 
occur in open multiprobe mesoscopic conductors. 
We have illustrated  that a small
absorption or a small emission of particles 
leads naturally to the appearance of these density 
of states. We have shown that the 
transmission probabilities through weakly coupled contacts, like the 
STM, is related to these densities. We have shown that a weakly
coupled voltage probe, describing inelastic scattering or a dephasing
process can be described in terms of these densities. We have given 
a formulation of the Friedel sum rule which holds even when the conditions
for the WKB approximation do not apply. 
We have also pointed 
out that the ac-conductance of a mesoscopic conductor at small frequencies
can be formulated in terms of these densities of states. Furthermore, we have 
indicated that it is useful to consider also the off-diagonal elements of 
a partial density of states matrix since this permits a description 
of fluctuation processes. This consideration has led us to a 
local Wigner-Smith life-time matrix. We have emphasized that for many 
problems in electrical conductors, it is necessary to 
consider the effect of screening. We have treated screening 
in a simple one-loop random phase approximation. 

The partial density of states which we have used here provide 
a classification with respect to the contact (and or channel) index
of the scattering matrix. In problems with high spatial symmetry \cite{ZUEL}
we can envision also a classification with respect to the local
momentum. Interest in spin transport will require 
spin resolved partial density of states. In hybrid normal 
superconducting systems \cite{SUP1} a resolution of both electron 
and hole density of states is conceptually useful \cite{TG00,SP01}.

The description of electrical conduction processes in terms 
of transmission and reflection probabilities has become very 
well known. The fact that we can also express densities 
with the help of scattering matrices makes the scattering approach 
applicable not only to dc-transport but to a much wider range 
of electrical transport phenomena. 

\section*{Acknowledgement}
This paper is in honor of Professor Dr. N. Kumar on the 
occasion of his 60th birthday. 
This work was supported by the Swiss National Science Foundation.

\end{multicols}

\begin{thebibliography}{90}
\addcontentsline{toc}{section}{References}
%
\bibitem{ABG}     I.L. Aleiner, P.W. Brouwer, L.I. Glazman, 
                  Phys. Rep. (unpublished). 
                  cond-mat/0103008

\bibitem{GH}      G. Hackenbroich, 
                  Phys. Rep. {\bf 343}, 463 (2001). 
%
\bibitem{BB}      Ya. M. Blanter and M. B\"{u}ttiker, 
                  Phys. Rep. {\bf 336}, 1 (2000).

                  
                  
%
\bibitem{BTP2}    M. B\"{u}ttiker, H. Thomas, and A. Pr\^etre,
                  Z. Phys. B {\bf 94}, 133 (1994).  
%
\bibitem{MB93}    M. B\"{u}ttiker,
                  J. Phys.: Condensed Matter {\bf 5}, 9361 (1993).


\bibitem{MATH}    M. B\"uttiker, J. Math. Phys. {\bf 37}, 4793 (1996).

\bibitem{LAND}    R. Landauer, Phil. Mag. {\bf 21}, 863 (1970); 
                  Z. Phys. B {\bf 68}, 217 (1987). 

%
\bibitem{IMRY}    Y. Imry, in {\it Directions in Condensed Matter Physics},
                  edited by G. Grinstein  and G. Mazenko, 
                  (World Scientific Singapore, 1986). p. 101. 
                  
\bibitem{MB86}    M. B\"{u}ttiker, Phys. Rev. Lett. {\bf 57}, 1761 (1986).                 
                   
\bibitem{GRAM}    T. Gramespacher and M. B\"uttiker, 
                  Phys. Rev. B {\bf 56}, 13026 (1997). 
                   
\bibitem{GASP}    V. Gasparian, T. Christen, and M. B\"{u}ttiker, 
                  Phys. Rev. A {\bf 54}, 4022 (1996). 


%
\bibitem{ZHAO}    X. Zhao, J. Phys. Cond. Matter, {\bf 12}, 4053 (2000).

%
\bibitem{DEJE}    Tiago De Jesus, Hong Guo, and Jian Wang, 
                  Phys. Rev. B {\bf 62}, 10774 (2000).    
                  
%
\bibitem{SMITH}   F. T. Smith, Phys.\ Rev.\ {\bf 118} 349 (1960). 

%
\bibitem{DASH}    R. Dashen, S. -k Ma and H. J. Bernstein, Phys. 
                  Rev. {\bf 187}, 345 (1969). 
                                
%
\bibitem{STAF}    F. Kassubek, C. A. Stafford and H. Grabert, 
                  Phys. Rev. B {\bf 59}, 7560 (1999). 

\bibitem{LARMOR}  M. B\"uttiker, Phys. Rev. B {\bf 27}, 6178 (1983).

\bibitem{BL82}    M. B\"{u}ttiker and R. Landauer, Phys. Rev. Lett.
                  {\bf 49}, 1739 (1982); Physica  Scripta {\bf 32},
                  429-434, (1985). 
                                
\bibitem{LEAE1}   C. R. Leavens and G. C. Aers, 
                  Solid State Commun. {\bf 63}, 1107 (1989).

\bibitem{LEAE2}   C. R. Leavens and G. C. Aers
                  Phys. Rev. B {\bf 40}, 5387 (1989).                   

\bibitem{MB01}    M. B\"uttiker, in "Time in Quantum Mechanics", 
                  edited by J. G. Muga, R. Sala Mayato and I. L. Egusquiza,
                  (unpublished). quant-ph/0103164 
                                    
\bibitem{MB90}    M. B\"{u}ttiker, in "Electronic Properties of Multilayers and low
                  Dimensional Semiconductors",
                  edited by J. M. Chamberlain, L. Eaves, and J. C. Portal,
                  (Plenum, New York, 1990). p. 297-315.
                  
\bibitem{KUMAR}   A. Rubio and N. Kumar,
                  Phys. Rev. B{\bf 47}, 2420 (1993);  
                  S. Anantha Ramakrishna and N. Kumar,  
                  Phys. Rev. B{\bf 61}, 3163 (2000). 
                 

\bibitem{CWJB}    C.W.J. Beenakker, in: Photonic Crystals and Light 
                  Localization in the 21st Century, edited by C.M. Soukoulis,
                  NATO Science Series C563 (Kluwer, Dordrecht, 2001): 
                  pp. 489. cond-mat/0009061

     
\bibitem{STM}     G. Binnig and H. Rohrer, 
                  Helv. Phys. Acta {\bf 55}, 726 (1982); 
                  J. Tersoff and D. R. Hamann, 
                  Phys. Rev. B {\bf 31}, 805 (1985). 
                  

\bibitem{STM2}     H. Watanabe, C. Manabe, T. Shigematsu, and M. Shimizu,
                   Appl. Phys. Lett. {\bf 78}, 2928 (2001). 

\bibitem{STM3}     P. Bxggild, T. M. Hansen, O. Kuhn, and F. Grey, 
                   Review of Scientific Instruments {\bf 71}, 2781 (2000). 

\bibitem{ENGQ}     H.-L.\ Engquist and P.\ W.\ Anderson, 
                   Phys.\ Rev.\ B {\bf 24}, 1151 (1981)                        

\bibitem{MB89}     M. B\"{u}ttiker, Phys. Rev. B {\bf 40}, 3409 (1989).
 

\bibitem{MB88}     M. B\"{u}ttiker,
                   IBM J. Res. Develop. {\bf 32}, 63 (1988).
              
\bibitem{PBCB}     P. W. Brouwer and C. W. J. Beenakker, 
                   Phys. Rev. B {\bf 55}, 4695 (1997). 
                  
                  
\bibitem{JAY}      T. P. Pareek, Sandeep K. Joshi, A. M. Jayannavar, 
                   Phys. Rev. B {\bf57}, 8809 (1998). 
                  
\bibitem{DEO}      P. Singha Deo, Swarnali Bandopadhyay, Sourin Das,
                   (unpublished). cond-mat/0102095; 
                   P. Singha Deo, (unpublished). cond-mat/0005123 

\bibitem{MBTC}     M. B\"{u}ttiker and T. Christen, in 
                   "Mesoscopic Electron Transport", 
                   NATO Advanced Study Institute, Series E: Applied Science, 
                   edited by L. L.  Sohn, L. P. Kouwenhoven and G. Schoen,
                   (Kluwer Academic Publishers, Dordrecht, 1997). 
                   Vol. 345. p. 259. 
                   cond-mat/9610025  
                   
\bibitem{ANKA}     M. B\"{u}ttiker, 
                   in "Quantum Mesoscopic Phenomena and Mesoscopic Devices", 
                   edited by I. O. Kulik and R. Ellialtioglu, (Kluwer, 
                   Academic Publishers, Dordrecht, 2000). Vol. 559, p. 211.
                   cond-mat/9911188  
                   
                   
                   
\bibitem{BR1}      P.W. Brouwer, Phys. Rev. B {\bf 58}, R10135 (1998)                   

\bibitem{AVRON}    J. E. Avron, A. Elgart, G. M. Graf, and L. Sadun, 
                   Phys. Rev. B {\bf 62}, R10618 (2000). 
                   

\bibitem{BR2}      M. L. Polianski and P. W. Brouwer, Phys. Rev. B{\bf 64}, 
                   075304 (2001).   
                   
                    
                   
\bibitem{LEVIN2}   Y. Levinson, Phys. Rev. B {\bf 61}, 4748 (2000).


\bibitem{ENTIN}    O. Entin-Wohlman, Y. Levinson, P. Wolfle, (unpublished).
                   cond-mat/0104408 
     

\bibitem{SCHOM}    H. Schomerus, M. Titov, P. W. Brouwer, C. W. J. Beenakker,
                   (unpublished). cond-mat/0107383 


               
\bibitem{PLB}      M. H. Pedersen, S. A. van Langen and M. B\"{u}ttiker,
                   Phys. Rev. B {\bf 57}, 1838 (1998). 
                   

\bibitem{BUKS1}    E. Buks, R. Schuster, M. Heiblum, D. Mahalu and V. Umansky,
                   Nature {\bf 391}, 871 (1998);                    
                   D. Sprinzak, E. Buks, M. Heiblum and H. Shtrikman,
                   Phys. Rev. Lett. {\bf 84}, 5820 (2000). 

\bibitem{MBAM}     
                   M. B\"uttiker and A. M. Martin, 
                   Phys. Rev. B{\bf 61}, 2737 (2000).
                   
   
\bibitem{LEVIN} 
                   Y. B. Levinson, Europhys. Lett. {\bf 39}, 299 (1997);
                   L. Stodolsky, Phys. Lett. B {\bf 459}, 193 (1999).                                                            



                   
\bibitem{AMMB}     A. M. Martin and M. B\"uttiker, 
                   Phys. Rev. Lett. {\bf 84}, 3386 (2000).                   
                   
                                                                          
\bibitem{SOMMERS}  Y.~V.\ Fyodorov and
                   H.~J.\ Sommers, Phys. Rev. Lett. {\bf 76}, 4709 (1996);
                   V.~A.\ Gopar, P.~A.\ Mello, and M.\
                   B\"{u}ttiker, Phys.\ Rev.\ Lett.\ {\bf 77}, 3005 (1996);
                   P.~W.\ Brouwer, K.~M.\ Frahm, and C.~W.~J.\ Beenakker,
                   Phys.\ Rev.\ Lett.\ {\bf 78}, 4737 (1997);
                   H.-J. Sommers, D. V. Savin, and V. V. Sokolov, 
                   Phys. Rev. Lett. {\bf 87}, 094101 (2001) 
                                  

\bibitem{OLDKU}    A. M. Jayannavar, G. V. Vijayagovindan, and N. Kumar, 
                   Z. Phys. B {\bf 75}, 77 (1989).                   

\bibitem{TEXIER}   C. Texier and A. Comtet, Phys. Rev. Lett. {\bf 82}, 4220 (1999). 
                   

\bibitem{ZUEL}     D. Boese, M. Governale, A. Rosch, and U. Z\"ulicke,                 
                   Phys. Rev. B {\bf 64}, 085315 (2001). 
                  
                  
\bibitem{SUP1}     J. Wang, Y. Wei, H. Guo, Q.-f. Sun, and T.-h. Lin
                   Phys. Rev. B {\bf 64}, 104508 (2001)


\bibitem{TG00}     T. Gramespacher and M. B\"uttiker, 
                   Phys. Rev. B {\bf 61}, 8125-8132 (2000).               
               
               
\bibitem{SP01}     S. Pilgram, H. Schomerus,  A. M. Martin, M. B\"uttiker, 
                   (unpublished). cond-mat/0107048 
                
\end{thebibliography}
\end{document}